# Impact of Product Complexity on Actual Effort in Software Developments: An Empirical Investigation

Zheng Li
School of Computer Science
NICTA and ANU
Canberra, Australia
Zheng.Li@nicta.com.au

Liam O'Brien
ICT Innovation and Services
Geoscience Australia
Canberra, Australia
Liamob99@hotmail.com

Ye Yang
Institute of Software
Chinese Academy of Sciences
Beijing, China
Yangye@nfs.iscas.ac.cn

*Abstract—* *[Background:]* Software effort prediction methods and models typically assume positive correlation between software product complexity and development effort. However, conflicting observations, i.e. negative correlation between product complexity and actual effort, have been witnessed from our experience with the COCOMO81 dataset. *[Aim:]* Given our doubt about whether the observed phenomenon is a coincidence, this study tries to investigate if an increase in product complexity can result in the abovementioned counter-intuitive trend in software development projects. *[Method:]* A modified association rule mining approach is applied to the transformed COCOMO81 dataset. To reduce noise of analysis, this approach uses a constant antecedent (Complexity increases while Effort decreases) to mine potential consequents with pruning. *[Results:]* The experiment has respectively mined four, five, and seven association rules from the general, embedded, and organic projects data. The consequents of the mined rules suggested two main aspects, namely human capability and product scale, to be particularly concerned in this study. *[Conclusions:]* The negative correlation between complexity and effort is not a coincidence under particular conditions. In a software project, interactions between product complexity and other factors, such as Programmer Capability and Analyst Capability, can inevitably play a "friction" role in weakening the practical influences of product complexity on actual development effort.

*Keywords – Product Complexity; Software Development; Software Effort Estimation; Empirical Software Engineering*

I. INTRODUCTION

Complexity has been recognized as being an essential property and intrinsic characteristic of software products [3, 21, 22], while product complexity has been viewed as the main source of the complexity of corresponding software projects [23, 24] and as a significant determinant of software development effort [4, 24, 25]. Moreover, software effort prediction methods and models typically assume positive correlation between product complexity and development effort [7, 18]. However, when trying to employ such a positive correlation as a valid assertion [26, 27] for software effort judgment, there is still a lack of empirical investigations as solid evidence. Therefore, we proposed to use empirical studies to reinforce the published knowledge.

For the convenience of identifying data to do the empirical investigation, we naturally adopted the well-known and well-documented COCOMO81 dataset [12]. After observing the data of 63 projects used in the COCOMO model, however, we could not find the positive correlation between software product complexity and development effort. On the contrary, the initial analysis showed a frequent trend of negative correlation between product complexity and actual effort. Roughly speaking, this phenomenon may not be surprising, because there are many effort factors interacting with each other during software developments. Nevertheless, according to the parsimony principle, "the mission of science is to come up with a short list of the most important factors; it is unacceptable to say 'everything depends on everything else'" [29]. Moreover, does such a phenomenon randomly happen? Or is there any rule or principle behind this? We doubt that the aforementioned trend is a coincidental phenomenon due to the random wax and wane of different effort drivers. Inspired by Lenz's Law [16] about the opposite directions of an induced electromotive force and the produced current in electromagnetism, we propose a set of research questions around the idea that an increase in product complexity would result in interactions with other factors that could weaken and even overwhelmingly weaken the complexity's influence on actual effort in software projects. In fact, it has been revealed that effort factors in real projects are hardly independent of each other [28]. A causal relationship may exist between different factors, i.e., "a factor's change leads to a change to a related factor". Therefore, the empirical study for answering those research questions can be also viewed as a further work on factor dependencies [28], which specifically investigates what factors may causally depend on product complexity in software projects.

In detail, we designed an experiment using the modified Apriori algorithm [9] to mine product complexity-related association rules only from data with the aforementioned trend. The data mining result suggested a set of rules that could act as possible answers to the abovementioned research questions. Through the mined rules, first of all, human capability (in terms of programmer capability and/or analyst capability) and product scale (in terms of product size and/or database size) were identified as two main factors that could overwhelmingly impact the influence of product complexity on actual effort, i.e.:

- Employing people with higher capabilities is a frequent condition that brings the negative

correlation between software product complexity and actual development effort.
- Developing software with smaller scale is a frequent condition that brings the negative correlation between software product complexity and actual development effort.

Moreover, by using the six-point scale of effort drivers rated in the COCOMO81 dataset, we further investigated the extent to which these two factors could overwhelm product complexity with respect to their influences on actual effort. Interestingly, the correlation between product complexity and actual effort fluctuates significantly across the six complexity scales, though different types of software projects have different correlation fluctuations. To establish an explanation chain to support the initial answers, at last, we also reported our analyses together with hypotheses to try to reveal why those frequent conditions could happen, such as:
- Human capability increases discretely, while it is unavoidable to build software development team with more than enough capabilities in particular changing intervals of product complexity.
- In a comparable context, higher complexity software products (modules) may imply smaller scale of the products (modules).

Note that our work does not deny the previous studies in [5, 6, 7, 18]. We believe that the positive correlation between complexity and effort is an ideal situation without considering the influences of other factors, while the negative correlation between product complexity and actual effort is a frequent phenomenon in practice. In particular, this work suggests that, when concerning or judging the influence of product complexity on software development effort, the corresponding human capability and product scale should be further and particularly considered.

The remainder of this paper is organized as follows. Section II introduces the inconsistency between the practical data and theoretical discussion about the complexity-effort correlation in software projects. Section III briefly describes the design of experiment used in our investigation. Section IV elaborates the experimental results and the corresponding analyses. Conclusions and some future work are summarized in Section V.

## II. INITIAL IMPRESSION: FROM LITERATURE TO PRACTICAL DATA

When it comes to the research in the relationship between software product complexity and development effort, we can unfold study along two ways: one is to get familiar with the relevant knowledge from the literature, while the other is to empirically investigate real data of past software projects. Here we both reviewed literature and observed real data to achieve an initial impression about the impact of product complexity on actual effort of software projects.

### A. Philosophy from the Literature

Although Product and Project are separate concepts, there is a close relationship between them. For design-and-implementation projects such as goods manufacture, building construction, or software development, product is the physical achievement of a project, and the major source of project complexity is the complexity of the product to be delivered [1]. Moreover, in a design-and-implementation project, even the complexity level of the manufacturing system is often determined by the complexity of the manufactured product itself [2]. Therefore, the product complexity plays a significant role in the overall complexity in a project.

In the software economics field, complexity is also viewed as an inherent property of the functional requirements of a software product, which cannot be reduced or simplified beyond a certain threshold [3, 21, 22]. Similarly, product complexity has been viewed as the main source of the complexity of the corresponding software projects [23, 24], and also been claimed to be a significant and non-negligible factor that influences the effort of software development and maintenance [4, 24, 25]. As such, a positive correlation between software complexity and development effort exists in many estimation models: "a more complex piece of software will generally require greater effort in development than a less complex counterpart" [18]. For example, in the COCOMO model, product complexity is treated as one of the 15 independent variables on the dependent variable – development effort [12]. The hockey stick function [7] vividly and qualitatively describes the abovementioned relationship when people are dealing with things, as illustrated in Figure 1. The amount of required effort may dramatically increase when the corresponding things exceed a certain level of complexity.

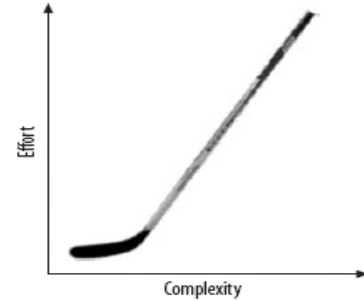

Figure 1. The hockey stick function.

In common sense, such assertions about impact of product complexity on development effort can be intuitively supported by mental reasons: The more complexity involved in a software product, the more difficulty the designers or engineers have to understand the development process and thus the product itself [5], and hence the greater mental effort people have to exert to solve the complexity [6].

### B. Observation of the Real Data

To empirically investigate the impact of product complexity on actual effort, the first and the most intuitive step is to observe real data. Here we employ the well-known and well-documented COCOMO81 dataset that comprises 63 real software projects [12]. Each software project in COCOMO81 uses a six-point scale to rate the project's 15 effort drivers including the Product Complexity: Very Low (VL), Low (L), Nominal (N), High (H), Very High (VH), and Extra High (XH). These rating values can be used conveniently for qualitative comparison between projects with respect to particular effort drivers. Therefore, inspired by the aforementioned hockey stick function, we can qualitatively observe the correlation between product complexity and actual effort by comparing the 63 projects with each other. Without considering the project development modes, the data of those 63 projects can be transformed into a qualitative comparison table with 1953 (=63×(63−1)÷2) records by using the equation (1). The comparison table is similar to Appendix I while eliminating the DEV_MODE column.

$$Attr_{i(Pj)} - Attr_{i(Pk)} = \begin{cases} + & \text{if } Attr_{i(Pj)} > Attr_{i(Pk)} \\ - & \text{if } Attr_{i(Pj)} < Attr_{i(Pk)} \\ 0 & \text{if } Attr_{i(Pj)} = Attr_{i(Pk)} \end{cases} \quad (1)$$

In equation (1), the $Attr_{i(Pj)}$ represents the $i$th attribute of the $j$th project $P_j$ in COCOMO81. Note that the attribute here refers not only to those 15 effort drivers that have discrete values, but also to the other two features of a project: Lines of Code and Actual Effort that have numeric values. When implementing comparison between two projects, the operations >, <, and = denote "higher than", "lower than", and "equal to" respectively for discrete values, while "bigger than", "smaller than", and "equal to" for numeric values. In particular, for our convenience of observation, we switch the sequence of two projects if the comparison result of product complexity is "−". In other words, the comparison result of product complexity in the qualitative comparison table can only be "+" or "0". Moreover, considering we are now focusing on the "correlation" that implies changing of the corresponding effort drivers, the records with value "0" in the column CPLX (CPLX stands for Product Complexity in COCOMO81) can be further pruned from the comparison table, because there is no change between two projects with respect to product complexity. In fact, it has already been clarified that project data having the same value of a factor would not support the conclusion about the interaction of that factor with other variables [8].

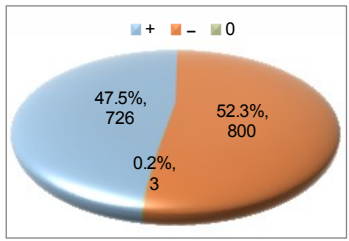

Figure 2. Distribution of the changes in actual effort when product complexity increases.

After pruning the comparison table, the remaining records can then represent the changes in the other project variables when, and only when, the product complexity increases. Therefore, we can conveniently observe the intuitive impact of increased product complexity on the other project features. When observing the correlation between the projects' actual effort and product complexity, surprisingly, more than half of the comparison results show a decrease in the actual effort in association with an increase in the product complexity, as illustrated in Figure 2.

In addition to the observation on overall project data, we also distinguish between different project-development modes. Project development in the COCOMO81 dataset belongs to one of three different modes: Organic, Semidetached, and Embedded [12]. When transforming the original data of those 63 projects into development mode-aware comparison table, we define that the data can be compared only between projects with the same development mode. Partial transformation results have been presented in Appendix I. Similarly, we also make sure the comparison result of product complexity can only be "+" or "0", and the records with value "0" in the column CPLX have been pruned from Appendix I. Given that the final transformation result covers only scenario of increasing product complexity, the distribution of the changes in actual effort under different development modes can be seen in Figure 3. Surprisingly again, the more complicated the development mode, the stronger the trend that there is a negative correlation between actual project effort and software product complexity.

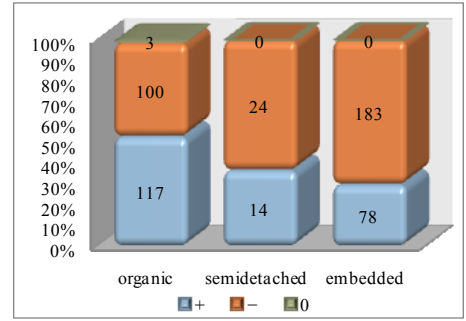

Figure 3. Distribution of the changes in actual effort when product complexity increases under different software development modes.

To summarize, apparently, practices of software development are not completely consistent with, and even show opposite trend to, the current theory about the relationship between product complexity and development effort. As such, it could be unreliable to apply the existing complexity-related knowledge to effort estimation when implementing software projects. To better understand the practical process of software development, it is necessary to investigate what makes practice not always obey the relevant theory. We do this through an empirical investigation outlined in the next section.

### III. DESIGN OF THE EMPIRICAL INVESTIGATION

The empirical investigation can be roughly divided into three steps, as listed below:
1. Define research questions.
2. Determine experimental method and implement experiments.
3. Analyse experimental results and answer research questions.

This section only introduces the first two investigation steps, while leaving the experimental result analysis and discussion to be elaborated in the next section.

## A. Research Questions of the Investigation

According to the previous review and observation, the essence of research questions to be defined is the negative correlation between software product complexity and actual development effort. Considering complexity is an inherent property of a software product, and inspired by Lenz's Law [16] in electromagnetism, "a current produced by an induced electromotive force (emf) moves in a direction so that its magnetic field opposes the original change in flux", we propose a root research question targeting the abovementioned counter-intuitive phenomenon:

**Q0:** Can an increase in product complexity trigger changes in other factors that weaken product complexity's influence on actual effort when developing software products?

In fact, effort factors of software development are not ideally independent of each other in practice [28]. It is possible that changing a factor results in interactions with other factors. In particular, this study focuses on the potentially causal relationship between product complexity and its related factors. Therefore, we first take into account the factors affected by the increase in product complexity, which results in two research questions:

**Q1:** Which effort factors' changes, triggered by the increase in product complexity, intend to weaken product complexity's influence on actual effort when developing software products?

**Q1':** Which effort factors' changes, triggered by the increase in product complexity, do NOT intend to weaken product complexity's influence on actual effort when developing software products?

Note that the Q1' is a side research question, and the potential answers to Q1' are not useful for this investigation. Given the answers to Q1, the negative correlation between complexity and effort cannot yet come into existence unless the triggered influence on effort oppositely overwhelms that of product complexity. Meanwhile, intuitively, the actual effort cannot be always in a decreasing trend while the product complexity keeps increasing. Therefore, it is also necessary to identify the abovementioned overwhelming extent, which can be addressed in a new research question:

**Q2:** To what extent can other factors overwhelm product complexity in terms of their opposite influences on actual effort when developing software products?

The study to answer the research questions Q1 and Q2 can be viewed as phenomenon identification. To better support Q0, the backend reasons of why such a phenomenon exists should be further revealed, as represented in Q3:

**Q3:** Why can an increase in product complexity trigger changes in other factors that weaken product complexity's influence on actual effort when developing software products?

To sum up, the root research question Q0 can be answered by answering questions Q1, Q2 and Q3.

## B. Experimental Method of the Investigation

To investigate what makes practice not always obey the existing knowledge in this case, naturally, we can focus only on the subset of project comparison data that shows negative correlation between actual effort and product complexity. Considering that we are to identify the underlying regularities composed of project attributes, a suitable investigation method could be data mining for association rules. In particular, here the antecedent of association rules has been pre-assigned as "CPLX='+' & ACTUAL='−'".

In data mining, association rules are derived by finding frequent item sets from a dataset [9]. Once frequent item sets are obtained, corresponding association rules can be straightforwardly generated with a level of confidence larger than or equal to a predefined minimum confidence [10]. The frequency of an item set, also called coverage, is the proportion of instances that covers the item set. The confidence of an association rule, also called accuracy, is the ratio of the number of instances that it predicts correctly to the number of instances to which it applies.

$$F = \frac{X \times 100\%}{N_{all}}, X = \sum_{i=1}^{N_{all}} \begin{cases} 1 & if\ i^{th}\ instance\ covers\ item\ set \\ 0 & otherwise \end{cases}. \quad (2)$$

Equation (2) shows the calculation of frequency $F$ of an item set: $X$ represents the number of instances covering the item set, and $N_{all}$ represents the number of all the instances in the original dataset. In this investigation, $N_{all}$ refers to the size of the previously mentioned subset on which we are focusing. With regard to the minimum frequency, we can obtain inspiration from the concept Majority Opinion that has been systematically explored in social psychology [14]. Past psychology studies reveal a pervasive tendency for individuals to follow majority positions in society. Similarly, we can follow the majority opinion and accept a coverage as long as it comprises more than half of the instances in the subset. Therefore, the minimum frequency in this association rule mining can be set as, but not include, 50%, i.e., $F > 50\%$.

$$A = \frac{Y \times 100\%}{N_{apply}}, Y = \sum_{j=1}^{N_{apply}} \begin{cases} 1 & if\ j^{th}\ instance\ obeys\ the\ rule \\ 0 & otherwise \end{cases}. \quad (3)$$

Equation (3) shows the calculation of accuracy $A$ of an association rule: $Y$ represents the number of correctly predicted instances by the rule, and $N_{apply}$ represents the number of instances to which the rule applies. Note that in this investigation $N_{apply}$ varies with changing item sets due to the pruning, which is further explained in the analysis of the rule mining algorithm below. When it comes to setting a threshold as minimum confidence, we try to borrow ideas from the performance assessment for effort estimation models. When assessing effort estimation models, in general, both Percentage Relative Error Deviation within x (PRED(x)) and Mean Magnitude Relative Error (MMRE) adopted 25% as measurement threshold [11, 13, 15]. Similarly, here we define that an association rule is acceptable if its incorrect predictions are less than 25% of applicable instances in a dataset. In other words, the minimum confidence for generating association rules can be set as, but not include, 75%, i.e., *A* > *75%*.

In particular, we have $X = Y$ when using equations (2) and (3) to calculate frequency and accuracy. On the one hand, since an association rule derived from an item set necessarily covers the item set, the instances correctly predicted by the rule then also cover the item set. On the other hand, only the instances that cover the item set obey the derived association rules. Therefore, the $Y$ for an associate rule is equal to the $X$ for the corresponding item set.

TABLE I. MODIFIED APRIORI ALGORITHM

```
ArrayList MiningAssociationRules ( string[,] comparisonTable )
{
    ArrayList AR = new ArrayList(); //To save association rules.

    //Prune table for pre-assigned antesedent.
    string[,] CT = comparisonTable;
    Prune rows from CT if CPLX != "+" or ACTUAL != "−";

    for ( int i = 1; i <= number of project features −2; i++ )
    {
        foreach ( combination of project features &&
            number of items in conbination = = i &&
            items in combination do no include CPLX or ACTUAL)
        {
            //Prune table for candidate consequent.
            string[,] T = CT;
            Prune rows from T if items in combination have value
                "0" in those rows;

            foreach ( cobination of values )
            {
                double appearance = number of  rows having
                                    combination of project features
                                    with combination of values in T;
                double applied = number of rows in T;
                double total = number of rows in pruned CT;
                if ( appearance / total  > 0.5 &&
                    appearance / applied > 0.75 )
                        AR.Add( "IF CPLX='+' and ACTUAL='−'
                            THEN " + combination of project
                            features with combination of values );
            }
        }
    }
    return AR;
}
```

After positioning the minimum frequency and confidence, we can use an Apriori-like algorithm [9] to mine association rules, as shown in Table I. Given the pre-assigned antecedent, this investigation needs only to derive different consequents from the dataset to build different association rules. Furthermore, to reduce the noise of analysis as explained previously [8], we prune the instances having items with value "0" if the items appear in a potential consequent. Note that the pruning makes the rule mining algorithm used in this investigation different from the classical Apriori algorithm that executes the rule-induction procedures for every possible combination of attributes, with every possible combination of values [9].

IV. RESULTS AND DISCUSSIONS

Following the sequence of previous observations on the COCOMO81 dataset, we apply the aforementioned algorithm to the transformed comparison data (see Appendix I) without and with distinguishing software development mode respectively. The derived association rules are correspondingly listed in Table II ~ Table IV.

*A. Discussion around Research Question Q1*

From the experiment without taking into account software development mode, we can achieve four association rules with consequents covering three project features: LOC, DATA, and PCAP, as shown in Table II.

TABLE II. ASSOCIATION RULES WITHOUT DISTINGUISHING PROJECT DEVELOPMENT MODES

| IF CPLX = "+" & ACTUAL = "−" THEN | | |
|---|---|---|
| ID | Consequent | Appearance / Total = Accuracy × 100% |
| 1 | LOC = "−" | 722/795=90.82% |
| 2 | PCAP = "+" | 487/611=79.71% |
| 3 | DATA = "−" | 511/621=82.29% |
| 4 | LOC = "−" & DATA = "−" | 479/617=77.63% |

In detail, LOC denotes the Product Size by using source lines of code; DATA refers to the Database Size that indicates the amount of data to be assembled and stored; PCAP represents the Programmer Capability including ability, efficiency, thoroughness, and communication/cooperation skills of developers who work together on a project. As such, those four rules can be summarized into a frequent phenomenon: When developing software projects in general, if product complexity increases while actual effort decreases, then the projects have programmers with higher capabilities while having software with smaller product size and/or database size. This phenomenon therefore reveals two possible answers to the research question Q1:

**A1 (to Q1):** For software projects, when product complexity increases, the actual effort can still decrease due to the increased capabilities of programmers.

**A2 (to Q1):** For software projects, when product complexity increases, the actual effort can still decrease due to the decreased product size or/and database size.

Similarly, we can respectively investigate the association rules generated from the experiments concerning different development modes. For software projects with the embedded mode, in addition to the aforementioned rules, another rule emerges with increasing the project feature RELY as its consequent, as shown in Table III. RELY refers to the Required Reliability that reflects how much a software product is expected to perform its intended functions during a specific period of time. Since the increase in RELY also intends to increase actual effort, the new rule then indicates a possible answer to the research question Q1':

**A3 (to Q1'): Software products with more complexity may also require more reliability.**

As mentioned earlier, such an answer does not suggest any hint about the negative correlation between product complexity and actual effort. Therefore, we do not give it more discussion in this paper.

TABLE III. ASSOCIATION RULES FOR EMBEDDED PROJECTS

| \multicolumn{3}{l}{IF CPLX = "+" and ACTUAL = "–" THEN} |||
|---|---|---|
| ID | Consequent | Appearance / Total = Accuracy × 100% |
| 1 | LOC = "–" | 169/183=92.35% |
| 2 | PCAP = "+" | 103/119=86.55% |
| 3 | DATA = "–" | 146/158=92.41% |
| 4 | RELY = "+" | 115/145=79.31% |
| 5 | LOC = "–" & DATA = "–" | 139/158=87.97% |

For software projects with the organic mode, in addition to the duplicate project features, ACAP and TURN appear in the generated association rules, as shown in Table IV.

TABLE IV. ASSOCIATION RULES FOR ORGANIC PROJECTS

| \multicolumn{3}{l}{IF CPLX = "+" and ACTUAL = "–" THEN} |||
|---|---|---|
| ID | Consequent | Appearance / Total = Accuracy × 100% |
| 1 | LOC = "–" | 75/99=75.76% |
| 2 | PCAP = "+" | 71/89=79.78% |
| 3 | ACAP = "+" | 76/86=88.37% |
| 4 | TURN = "–" | 60/72=83.33% |
| 5 | RELY = "+" | 63/77=81.82% |
| 6 | ACAP = "+" & PCAP = "+" | 63/80=78.75% |
| 7 | ACAP = "+" & RELY = "+" | 55/69=79.71% |

ACAP denotes Analyst Capability including ability, efficiency, thoroughness, and communication/cooperation skills of analysts as a team in a project. TURN represents Computer Turnaround Time that reflects the response time of development jobs handled by computers. Thus, the rules involving ACAP and TURN introduce a new frequent phenomenon: When developing organic-mode software projects, if product complexity increases while actual effort decreases, then the projects have analysts with higher capabilities while have computers with shorter response time to development jobs. From this phenomenon, we can also retrieve two additional answers to the research question Q1:

**A4 (to Q1): For organic-mode software projects, when product complexity increases, the actual effort can still decrease due to the increased capabilities of analysts.**

**A5 (to Q1): For organic-mode software projects, when product complexity increases, the actual effort can still decrease due to the decreased computer response time to development jobs.**

For software projects with the semidetached mode, unfortunately, there is a relative shortage of relevant data to support mining valid association rules. As a result, more and divergent rules are generated (63 in total), and some of them are related to irrational phenomena like decreasing the use of software tools to decrease effort. Therefore, in this investigation we ignore the data of semidetached-mode projects and do not elaborate/analyse the corresponding rules.

Overall, we can find that the **human capability** (in terms of the capabilities of programmers and/or analysts) and **product scale** (in terms of the size of product and/or database) are two main factors for the negative correlation between actual effort and product complexity in software developments. Nevertheless, as previously mentioned, it is impossible to infinitely increase product complexity without increasing actual effort by adjusting other factors. Therefore, it is worth investigating further to what extent people can spend less effort for a more complex software product.

*B. Discussion around Research Question Q2*

Considering the product complexity is rated on a six-point scale as mentioned earlier, we can naturally use the complexity scale to measure when the negative correlation between complexity and effort happens. In detail, we respectively investigate project data with consecutive CPLX rates following the method described in Subsection B of Section II. Through observing the consecutive changes in product complexity with the corresponding changes in actual effort, we can roughly and qualitatively identify the turning points where the direction of co-movement between effort and complexity overturns.

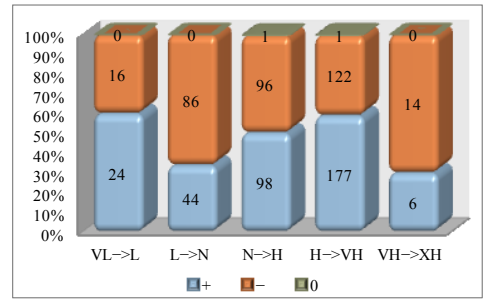

Figure 4. Distribution of the changes in actual effort when product complexity consecutively increases in general software projects.

When it comes to the observation on overall project data, the COCOMO81 dataset is first divided and packed into five subsets each of which comprises data with two consecutive product complexities; then, the aforementioned data transformation without distinguishing software development modes is implemented in every subset. Finally, we can observe distributions of the changes in actual effort at every consecutive increase in product complexity, as shown in Figure 4. Following the Majority Opinion [14] again, we can roughly claim the changing trends of actual effort when product complexity increases within every two consecutive scales. Interestingly, the actual effort shows a fluctuating change in association with the consecutive increase in product complexity.

For the convenience of discussion, we use a curve to qualitatively represent the co-movement between product complexity and actual effort for general software projects, as illustrated in Figure 5. Ignoring the two terminal scales VL and XH, there is one obverse turning point N and two reverse turning points L and VH in the co-movement curve. Note that H is only a bearing point where the changing trend of actual effort does not overturn. To help explain this curve, we can construct such an assumption: in general software development, product complexity plays a major role in driving actual effort if the complexity is lower than L; when product complexity is between L and N, other factors can overwhelm the product complexity in terms of their opposite influences on actual effort; from N to VH, product complexity dominates the increase in development effort again; while we can still adjust the other factors to decrease actual effort even if the product complexity is higher than VH. In brief, we can state that:

**A6 (to Q2): In general, other factors can overwhelm the product complexity in terms of their opposite influences on actual effort within two product complexity intervals: [L, N] and [VH, XH].**

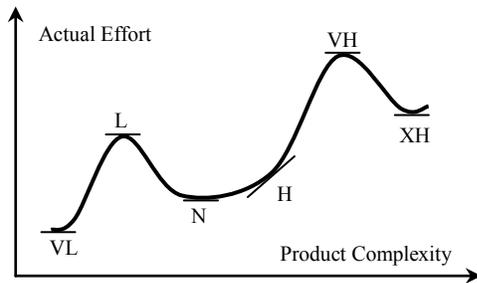

Figure 5. The qualitative curve of co-movement between actual effort and product complexity when developing software projects in general.

In succession, as mentioned previously, the project data under different development modes should be also analysed respectively, except for the semidetached projects. For organic projects, similarly, the result of data transformation and effort-changing distribution can be visualized as shown in Figure 6. The change in actual effort also shows an exceptional fluctuation when consecutively increasing product complexity.

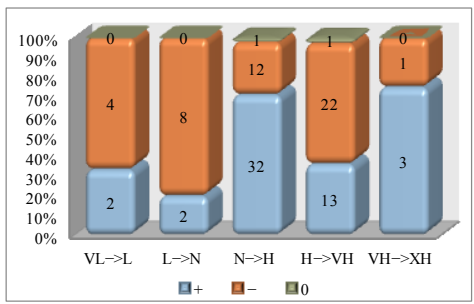

Figure 6. Distribution of the changes in actual effort when product complexity consecutively increases under organic development mode.

Accordingly, the complexity-effort co-movement curve for organic software projects can be qualitatively drawn as illustrated in Figure 7. In detail, the scale H of product complexity indicates a reverse turning point, while N and VH are obverse turning points. It is then possible to suppose that:

**A7 (to Q2): When developing organic software projects, other factors can overwhelm the product complexity in terms of their opposite influences on actual effort within two complexity intervals: [VL, N] and [H, VH].**

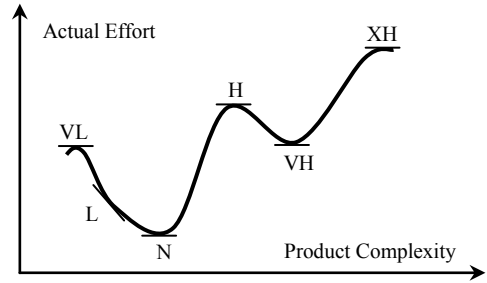

Figure 7. The qualitative curve of co-movement between actual effort and product complexity when developing organic software projects.

As for the embedded projects, the effort-changing distribution and the complexity-effort co-movement curve are displayed in Figure 8 and 9 respectively.

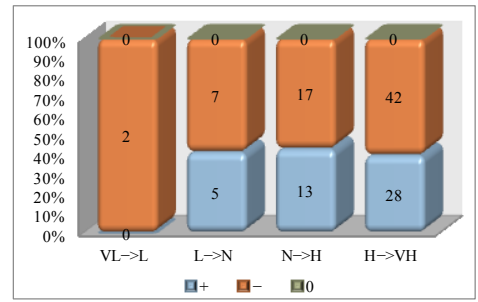

Figure 8. Distribution of the changes in actual effort when product complexity consecutively increases under embedded development mode.

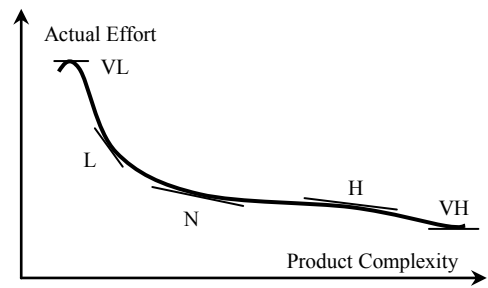

Figure 9. The qualitative curve of co-movement between actual effort and product complexity when developing embedded software projects.

Note that none of the embedded projects has XH complexity in the COCOMO81 dataset. These two figures reveal an even more extreme phenomenon in conflict with the hockey stick function (see Figure 1), and there is not any

turning point in the curve for embedded projects. We can therefore make another statement as:

**A8 (to Q2): When developing embedded software projects, other factors can overwhelm the product complexity in terms of their opposite influences on actual effort within the product complexity interval: [VL, VH]**.

*C. Discussion around Research Question Q3*

Since we have built up associations between the increase in product complexity and some other factors, we can intuitively assume that those factors are triggered by the increase in product complexity through the rules like "if CPLX = '+' (& ACTUAL = '−') then PCAP = '−'". Nevertheless, such assumptions could not make sense unless there are acceptable and supporting explanations. Therefore, with our main focus on the two factors mined previously, namely **human capability** and **product scale**, we try to identify possible reasons for that the increase in product complexity can trigger opposite influences on actual development effort.

*1) Reason Identification around Human Capability:*

When it comes to the human capability, it is clear that different people may have different capabilities. The same problem, while too complex to be solved for some people, can be easily done by others with higher capabilities. Therefore, a highly capable development team has to be built if a software product exceeds a certain level of complexity. Meanwhile, if holding the other aspects constant, it is also clear that the development team with higher than enough human capabilities would take less effort to complete a particular software project. In other words, even if product complexity increases, it is still possible to reduce actual effort by building up a development team with more than enough capabilities.

Furthermore, according to the aforementioned analyses related to Figure 5 and 7, we can find that other factors may repeatedly overwhelm product complexity in different phases. Given that only human capability and product size are discussed here, the only possible explanation is that:

**A9 (to Q3): Employing people with more than enough capabilities can be an unavoidable situation in particular product complexity intervals.**

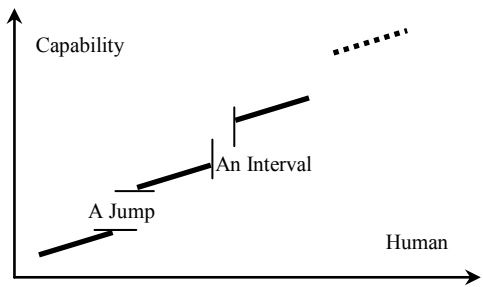

Figure 10. A rough and possible representation of human capability groups.

To assist the above answer, we hypothesize that:

**H1: Human capabilities could be scattered into several groups while intervals and/or jumps exist between different capability groups.**

In other words, unlike human experience that can be continuously measured by time, human capability could not always change continuously. By serially arranging people according to their capabilities, we can roughly represent grouped human capabilities as illustrated in Figure 10. Human capability changes in quantity in the same group, while it changes in quality when jumping into different groups.

*2) Reason Identification around Product Scale:*

As previously analysed, the reduced product scale can also decrease actual effort even when product complexity increases. Given this frequent phenomenon as revealed in the Table II ~ IV, a possible explanation is that:

**A10 (to Q3): Higher complexity software products (modules) may imply smaller scale of the products (modules).**

Considering that the complexity in a software product incurs difficulty for people to understand the development process and thus the product itself [5], less product scale that lowers the cognitive difficulty [19] may become a tradeoff between more complexity and successful implementation effort of a software project. Note that the statement of this answer may require a comparable context. For example, it is possible to construct an extremely complex and large software product by developing and composing smaller modules. In this case, it is fairer to compare between those product modules rather than between the final product and its modules. As such, this investigation also doubts the general claim about the positive correlation between the product complexity and software size [18, 20].

In addition, another possible reason is the adoption of tools or techniques in complex software developments. For example, by using a particular game development tool, people can create 64KB games, demos and screensavers [17]. Therefore, in the context of this discussion, we hypothesize that:

**H2: It is common and sometimes inevitable to employ tools/techniques to facilitate developing complex software products and as a result lessening the products' sizes.**

Note that this is a hypothesis instead of an answer because the aforementioned experiment did not mine the association between the product size (LOC) and the use of software tools (TOOL). However, this hypothesis and aforementioned human capability-related answers are consistent with previous relevant studies − "software development productivity still depends on the capabilities of people and tools involved" [28].

## V. CONCLUSIONS AND FUTURE WORK

By observing the COCOMO81 dataset, we found a clear inconsistency of practice and theory about the relationship between product complexity and actual effort of software projects. Admittedly, different effort drivers can have interactions between each other [28], which may imply complicated software development process and bring various phenomena. However, it is significant to investigate whether or not any regularity or rule exists behind a frequent phenomenon. In particular, we doubt that the aforementioned inconsistency is a coincidence due to the random wax and wane of different effort drivers. Inspired by Lenz's Law [16], we performed an empirical study to investigate the real effect of software product complexity on actual development effort. This investigation roughly and qualitatively verifies our consideration about that the increase in product complexity can trigger other factors that oppositely influence the actual effort in software projects. Nevertheless, we do not think our work denies the published knowledge. By analogy with the theoretical uniform motion without considering the friction in Physics, we can regard the existing studies [7, 18] as ideal approximation to the real relationship between product complexity and actual effort. In practice, the unavoidable interactions between effort drivers would play a "friction" role in weakening the effect of product complexity on actual effort of software projects.

Overall, this empirical investigation confirms that only concerning effort factors could be insufficient in the research into software development. Factor interactions and dependencies should be also taken into account when investigating or modeling software development practices [28]. In particular, human capability and product scale should be especially considered when estimating the influence of product complexity on software development effort in practice.

To establish an explanation chain to verify our original consideration, in fact, this study also poses some hypotheses that should be further tested and investigated. Therefore, our future work will be unfolded along two directions. Firstly, more experiments based on more datasets will be implemented to reinforce the complexity-effort study in this paper. Secondly, we will gradually start investigating the interactions between different effort drivers.

APPENDIX I. QUALITATIVE COMPARISON BETWEEN PROJECTS WITH THE SAME SOFTWARE DEVELOPMENT MODE

| ID | Projects Comparison | DEV_MODE | RELY | DATA | CPLX | TIME | STOR | VIRT | TURN | ACAP | AEXP | PCAP | VEXP | LEXP | MODP | TOOL | SCED | LOC | ACTUAL |
|---|---|---|---|---|---|---|---|---|---|---|---|---|---|---|---|---|---|---|---|
| 1 | #2−#1 | embedded | 0 | 0 | + | 0 | 0 | − | 0 | + | + | + | + | + | + | + | − | + | − |
| 2 | #8−#1 | embedded | + | − | + | + | + | + | − | + | + | + | − | − | + | 0 | − | − | − |
| 3 | #9−#1 | embedded | + | − | + | + | + | 0 | − | + | + | + | 0 | − | + | + | − | − | − |
| 4 | #10−#1 | embedded | + | − | + | + | + | − | 0 | + | + | + | + | 0 | + | + | − | − | − |
| 5 | #11−#1 | embedded | + | − | + | + | + | − | 0 | + | + | + | + | 0 | + | + | − | + | + |
| 6 | #12−#1 | embedded | + | − | + | + | 0 | − | − | + | + | + | + | + | + | + | − | − | − |
| 7 | #13−#1 | embedded | + | − | + | + | 0 | 0 | − | + | + | + | 0 | 0 | + | + | − | − | − |
| 8 | #15−#1 | embedded | + | − | + | + | 0 | 0 | − | + | 0 | + | − | − | + | + | − | − | − |
| 9 | #16−#1 | embedded | + | − | + | + | + | − | − | + | + | + | + | 0 | + | + | − | − | − |
| 10 | #17−#1 | embedded | + | − | + | + | + | − | − | + | + | + | + | 0 | + | + | − | − | − |
| 11 | #18−#1 | embedded | + | 0 | + | + | + | − | 0 | + | + | + | + | 0 | 0 | 0 | − | + | + |
| 12 | #19−#1 | embedded | + | − | + | + | + | − | − | + | + | + | + | 0 | + | + | − | + | + |
| 13 | #21−#1 | embedded | + | 0 | + | + | 0 | − | − | + | + | + | + | 0 | + | + | − | + | + |
| 14 | #22−#1 | embedded | + | − | + | + | 0 | − | − | + | + | + | + | 0 | + | + | − | + | − |
| 15 | #23−#1 | embedded | + | − | + | + | 0 | − | − | + | + | + | + | 0 | + | + | − | − | − |
| ... | ... ... | ... ... | ... ... | ... ... | ... ... | ... ... | ... ... | ... ... | ... ... | ... ... | ... ... | ... ... | ... ... | ... ... | ... ... | ... ... | ... ... | ... ... | ... ... |
| 491 | #60−#28 | organic | 0 | − | 0 | 0 | − | − | − | 0 | − | 0 | 0 | 0 | 0 | 0 | − | − | − |
| 492 | #28−#61 | organic | + | + | + | + | + | + | + | 0 | 0 | 0 | − | − | − | − | 0 | − | + |
| 493 | #30−#29 | semidetached | 0 | 0 | 0 | 0 | 0 | 0 | 0 | 0 | 0 | 0 | 0 | 0 | 0 | 0 | 0 | − | − |
| 494 | #29−#32 | semidetached | 0 | − | + | 0 | 0 | 0 | 0 | − | − | + | 0 | 0 | + | + | − | − | − |
| 495 | #29−#36 | semidetached | 0 | − | + | 0 | 0 | + | + | + | − | + | − | − | + | 0 | − | − | − |
| 496 | #29−#48 | semidetached | 0 | + | + | 0 | 0 | + | 0 | + | − | + | − | − | + | + | − | − | − |
| 497 | #29−#49 | semidetached | − | 0 | + | 0 | 0 | 0 | + | − | − | − | + | 0 | − | 0 | − | − | − |
| ... | ... ... | ... ... | ... ... | ... ... | ... ... | ... ... | ... ... | ... ... | ... ... | ... ... | ... ... | ... ... | ... ... | ... ... | ... ... | ... ... | ... ... | ... ... | ... ... |
| 706 | #58−#63 | embedded | + | 0 | + | + | + | 0 | + | 0 | 0 | + | + | + | − | 0 | 0 | + | + |
| 707 | #60−#59 | organic | + | 0 | + | 0 | 0 | 0 | + | + | − | + | − | − | − | − | − | − | − |
| 708 | #61−#59 | organic | 0 | 0 | 0 | − | − | − | 0 | + | 0 | + | + | 0 | + | 0 | 0 | + | − |
| 709 | #60−#61 | organic | + | 0 | + | + | + | + | + | 0 | − | 0 | − | − | − | − | − | − | + |